\documentclass[aps,twocolumn,10pt, 
superscriptaddress,
footinbib,
floatfix,
prx, longbibliography]{revtex4-2}

\usepackage{amsmath, amssymb, bm, braket, bbold, mathtools, graphicx, latexsym, epstopdf, tensor, hyperref, xfrac, comment, float, stackengine, bbm, orcidlink, xcolor, tabularx, multirow}
\usepackage[normalem]{ulem} 
\usepackage[ngerman,english]{babel}
\usepackage[utf8]{inputenc}
\usepackage[T1]{fontenc}
\usepackage[nointegrals]{wasysym}
\usepackage[export]{adjustbox}
\usepackage{natbib}

\begin{document}

\title{Detecting high-dimensional time-bin entanglement in fiber-loop systems}

\date{\today}

\author{Niklas Euler\,\orcidlink{0009-0009-2401-817X}}
\email[]{niklas.euler@uni-jena.de}
\affiliation{Institut für Festkörpertheorie und Optik, Friedrich-Schiller-Universität Jena, Max-Wien-Platz 1, 07743 Jena, Germany}
\affiliation{Physikalisches Institut, Ruprecht-Karls-Universität Heidelberg, Im Neuenheimer Feld 226, 69120 Heidelberg, Germany}

\author{Monika Monika\,\orcidlink{0000-0001-6986-9212}}
\affiliation{Institut für Festkörpertheorie und Optik, Friedrich-Schiller-Universität Jena, Max-Wien-Platz 1, 07743 Jena, Germany}
\affiliation{Fraunhofer-Institut für Angewandte Optik und Feinmechanik IOF, Albert-Einstein-Str. 7, 07745 Jena}

\author{Ulf Peschel\,\orcidlink{0000-0002-3980-9224}}

\author{Martin G\"{a}rttner\,\orcidlink{0000-0003-1914-7099}}
\email[]{martin.gaerttner@uni-jena.de}
\affiliation{Institut für Festkörpertheorie und Optik, Friedrich-Schiller-Universität Jena, Max-Wien-Platz 1, 07743 Jena, Germany}

\begin{abstract}
Many quantum communication protocols rely on the distribution of entanglement between the different participating parties.
One example is quantum key distribution (QKD), an application which has matured to commercial use in recent years.
However, difficulties remain, especially with noise resilience and channel capacity in long-distance communication.
One way to overcome these problems is to use high-dimensional entanglement, which has been shown to be more robust to noise and enables higher secret-key rates.
It is therefore important to have access to certifiable high-dimensional entanglement sources to confidently implement these advanced QKD protocols.
Here, we develop a method for certifying high-dimensional time-bin entanglement in fiber-loop systems.
In these systems, entanglement creation and detection can utilize the same physical components, and the number of time bins, and thus the entanglement dimension, can be adapted without making physical changes to the setup.
Our certification method builds on previous proposals for the certification of angular-momentum entanglement in photon pairs.
In particular, measurements in only two experimentally accessible bases are sufficient to obtain a lower bound on the entanglement dimension for both two- and multiphoton quantum states.
Numerical simulations show that the method is robust against typical experimental noise effects and works well even with limited measurement statistics, thus establishing time-bin encoded photons as a promising platform for high-dimensional quantum-communication protocols.
\end{abstract}
\maketitle  
\section{Introduction}
\label{sec:introduction}

The field of photonic quantum technologies has seen significant advances over the last decades \cite{obrien_photonic_2009, flamini_photonic_2018, slussarenko_photonic_2019, pelucchi_potential_2022}, enabling optical implementations in quantum simulation \cite{aspuru-guzik_photonic_2012, hartmann_quantum_2016}, quantum teleportation \cite{pirandola_advances_2015, llewellyn_chip--chip_2020}, and quantum computing applications \cite{knill_scheme_2001, nielsen_optical_2004, zhong_quantum_2020, bourassa_blueprint_2021, madsen_quantum_2022, couteau_applications_2023}, among others.
However, arguably the most advanced use case is quantum cryptography, especially quantum key distribution (QKD) \cite{bennett_quantum_2014, gisin_quantum_2002, gisin_quantum_2007,gisin_quantum_2010, lo_secure_2014, pirandola_advances_2020, luo_recent_2023}, with commercial solutions available for over two decades \cite{qkd_commercial_availability}.
This family of protocols encompasses a variety of ways to securely exchange secret keys to encrypt messages between distant parties.
Some QKD protocols utilize quantum entanglement as a resource, most notably those that implement device-independent QKD, the most secure branch of QKD \cite{acin_device-independent_2007, schwonnek_device-independent_2021, zapatero_advances_2023}.
This version of QKD makes no assumptions about the inner workings of the source of entanglement that the two parties share, and is even able to detect whether a malicious third party has taken over the source to manipulate the state preparation.
These abilities come with a price tag: In order to declare the channel secure with high success probability, the source needs to be of high quality and prepare the desired state with high fidelity.
Anoth pressing issue in practical applications is the high minimum photon detection efficiency $\nu\gtrsim 80\%$ required to guarantee protocol security \cite{brown_computing_2021, woodhead_device-independent_2021, zapatero_advances_2023}.
This is especially difficult to achieve for medium- to long-distance applications, as even state-of-the-art single-photon-detector technology cannot compensate the accumulating channel loss \cite{zapatero_long-distance_2019, esmaeil_zadeh_superconducting_2021}.

A possible approach to overcome this issue is the use of quantum states with high-dimensional entanglement.
The critical detection efficiency $\nu$ has been shown to decrease exponentially with the entanglement dimension $k$ of the shared photon pair \cite{massar_nonlocality_2002, miklin_exponentially_2022, zapatero_advances_2023, rivera-dean_device-independent_2024}.
Moreover, high-dimensional encoding in an entangled system has been shown to be more resilient to channel noise \cite{kaszlikowski_violations_2000, collins_bell_2002, durt_security_2003, durt_security_2004}.
Consequently, there is a demand for sources of high-quality, high-dimensional entanglement for QKD applications.

Fiber loop systems, where quantum information is encoded in discrete time intervals, also called time bins \cite{brendel_pulsed_1999}, have advantageous properties for realizing such sources. 
Key features are high resource efficiency, inherent phase stability, and fine-grained control in realizing interferometric schemes, called discrete time quantum walks (DTQWs), resulting in unprecedented scalability and a plethora of achievable interference patterns \cite{schreiber_photons_2010, nitsche_quantum_2016, he_time-bin-encoded_2017}.
Various loop topologies have been used to implement DTQWs \cite{schreiber_photons_2010, regensburger_photon_2011, schreiber_2d_2012, nitsche_quantum_2016} with the goal of studying boson sampling tasks \cite{motes_scalable_2014, boutari_large_2016, he_time-bin-encoded_2017} and even to probe topological quantum state properties \cite{nitsche_eigenvalue_2019}.
The use of time-bin encoding for quantum information processing and QKD applications has been a subject of ongoing research in recent years \cite{marcikic_distribution_2004, yu_two-photon_2015, boaron_simple_2018, flamini_photonic_2018, boaron_secure_2018, islam_scalable_2019, bouchard_measuring_2023, bouchard_programmable_2024}.
An important advantage is the insensitivity to polarization dispersion in optical fibers \cite{flamini_photonic_2018, fitzke_scalable_2022}, which makes active polarization stabilization necessary in polarization encoding, especially in long-distance settings \cite{xavier_experimental_2009, antonelli_sudden_2011, brodsky_loss_2011, ding_polarization_2017, craddock_automated_2024}.
This makes time-bin encoding a versatile and robust choice for quantum communication applications \cite{donohue_coherent_2013}.

In this work, we consider the architecture shown schematically in Fig.~\ref{fig:setup}. 
It features two coupled fiber loops that can be used both for generating high-dimensional entangled states and certifying them through optical path interference, providing a highly stable and compact experimental setup. 
We derive a certification protocol that combines time-of-arrival measurements taken with and without additional optical-path interference to obtain a lower bound on the fidelity to an adequately chosen reference state.
This bound serves as an entanglement-dimension witness for the prepared state.
Our protocol is widely applicable beyond the specific experimental setup considered here as its crucial ingredient, discrete quantum walks, can be realized on many physical multi-mode platforms.
We show through numerical simulations using realistic experimental and statistical error models that our method can certify high-dimensional entanglement with moderate experimental resource requirements, demonstrating its feasibility in the near future.
Our results highlight fiber-loop architectures as a competitive platform of resource-efficient high-dimensional entanglement preparation and certification with a high degree of programmability.

In the remainder of this paper, we first introduce the targeted fiber-loop architecture in Sec.~\ref{sec:setup} and describe the details of the certification method for time-bin-entangled photon pairs in Sec.~\ref{sec:results}.
Subsequently, we give the results of the aforementioned numerical simulations to showcase the performance of the methods in Sec.~\ref{sec:numerics}.
Finally, we also consider a multi-photon scenario in Sec.~\ref{sec:multi-spdc} before discussing our findings in Sec.~\ref{sec:discussion}.

\section{Capabilities of fiber-loop platforms}
\label{sec:setup}

\begin{figure}
    \centering
    \includegraphics[width=1\linewidth]{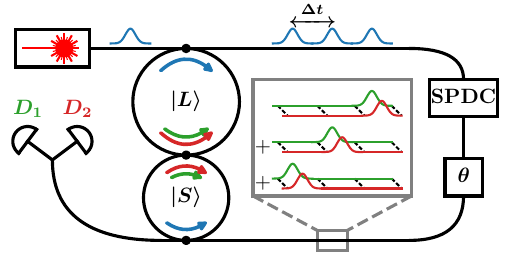}
    \caption{A sketch of the targeted experimental platform with illustrations of the different steps of its operation.
First, a single classical laser pulse (blue) is produced.
It is injected into the unbalanced loop system, where it is split into a series of pulses.
Through SPDC, one pair of signal and idler photons (green and red) is produced in a superposition in time over all possible source pulses (grey box).
A time-bin-dependent phase can be imprinted on this two-photon initial state.
The two SPDs $D_1$ and $D_2$ measure the joint two-photon arrival times.
Optionally, the quantum state can be injected into the loops to perform a DTQW before the detection.}
    \label{fig:setup}
\end{figure}

Our general strategy of entanglement certification builds on measurements in different bases.
We focus on the case where the corresponding basis transformations are accomplished using discrete quantum walks.
Contemporary fiber-loop platforms \cite{motes_scalable_2014, boutari_large_2016, nitsche_eigenvalue_2019, monika_quantum_2024} realize such quantum walks naturally and are thus a suitable platform for its experimental implementation.
For concreteness, we consider a temporal photonic lattice platform that has recently been realized \cite{monika_quantum_2024}.
A schematic overview of the key components of the setup is shown in Fig.~\ref{fig:setup}.
Here, we highlight some of the key working principles and capabilities of this kind of architecture; for an in-depth description of the experimental implementation, we refer to the original publication \cite{monika_quantum_2024}.

The system comprises two optical-fiber loops of unequal lengths, 100\,m and 120\,m, coupled by a dynamic fiber coupler, resulting in a photon round-trip time difference of $\Delta t = 100\,$ns.
Each loop is equipped with a gate to regulate the inflow and outflow of signals.
Light from the laser source is first injected into the long loop, then passes through a nonlinear optical medium and a phase modulator.
The generated time bins are either reinjected into the short loop before ultimately reaching two single-photon detectors (SPDs) or sent directly to the SPDs.

The protocol to generate a pair of photons with high-dimensional entanglement starts by injecting a classical laser pulse into the long loop; the fiber gates connecting the loops to the external fiber are shut afterward.
On every round trip, the pulse is split between the two loops.
Due to the difference $\Delta t$ in propagation time between the two loops, arbitrary sequences of equidistant pulses can be generated in the long loop by letting the pulse propagate for an appropriate number of round trips.
This method ensures high interpulse phase stability, as all $N$ pulses in the series originate from the same source pulse and have traveled the same distance in optical fiber at the point of interference.
The maximum number of pulses that can be accommodated is determined by the lengths of the loops and their length difference.
In the experiment reported in Ref.~\cite{monika_quantum_2024}, the loop configuration supported up to $N_{max}=4$ pulses.

Once the desired pulse sequence has been generated, it is ejected from the long loop and sent through the nonlinear optical medium to generate photon pairs via spontaneous parametric down-conversion (SPDC).
Postselection allows to project onto the case of one photon pair being created in one of the time bins.
In the resulting state,
\begin{align}
    \ket{\Psi(\boldsymbol{\alpha})}=\sum_{j=1}^N \alpha_j \ket{j}_S\otimes\ket{j}_I\eqqcolon\sum_{j=1}^N \alpha_j \ket{jj},\label{eq:state_out}
\end{align}
the signal and idler photon are being created in the same time bin $j$, in a superposition over all bins.
This state is already in a Schmidt-decomposed form, as the two photons are perfectly correlated in the time-bin basis.
In other words, the two photons of the output state share $N$-dimensional entanglement across their discrete time-bin degree of freedom.
The squared individual time-bin weight $|\alpha_j|^2$ corresponds to the amplitude of the $j$th classical pulse, which can be controlled through the dynamic coupling settings and by classical pulse modulation.
After the SPDC process, a time-bin-dependent phase can be imprinted on the state, resulting in
\begin{align}
\ket{\Psi(\boldsymbol{\alpha},\boldsymbol{\varphi})}=\sum_{j=1}^N \alpha_j e^{2i\varphi_j} \ket{jj}.\label{eq:phase-imprint}
\end{align}
Subsequently, the state populations can be measured directly in the time-bin basis by jointly recording the arrival times of the two photons at the SPDs.

Alternatively, it is also possible to feed the state into the loop system for a second time to perform a DTQW before measuring the time of arrival at the SPDs.
In particular, any given pulse can be delayed relative to the other pulses by a multiple of $\Delta t$ by temporarily coupling it into the long loop.
This ability sets fiber-loop platforms apart from time-bin platforms based on unbalanced Mach-Zehnder interferometers, which typically have a small number of fixed delay capabilities implemented on the hardware \cite{martin_quantifying_2017, madsen_quantum_2022, islam_provably_2017, bouchard_measuring_2023}.
Therefore, being able to perform arbitrary programmable quantum walks opens up further measurement capabilities to utilize in entanglement- and state certification.
For details on the underlying model of the central coupler and the resulting 1D DTQW, we refer to Appendix~\ref{app:DTQW}.

In the following section, we show the details of how path interference can be used to obtain useful bounds on high-dimensional entanglement in the prepared state.

\section{Certification scheme}
\label{sec:results}

As we showed in the previous section, the underlying fiber-loop architecture allows for a specific set of measurements, including arbitrary shifts and interference of time bins.
Here, we demonstrate how these can be used to measure a lower bound on the fidelity to a set of highly-entangled reference states,
\begin{align}
\ket{\Psi_{\mathrm{ref}}{}(\boldsymbol\lambda}) = \sum_{i=1}^N\lambda_i\ket{ii},\label{eq:reference_state}
\end{align}
which allows certifying high-dimensional entanglement \cite{sanpera_schmidt-number_2001, sentis_quantifying_2016, erker_quantifying_2017, bavaresco_measurements_2018, euler_detecting_2023}.

At its core, the method utilizes a family of upper bounds on the fidelity $F$ between the prepared state $\hat{\rho}$ and the reference state $\ket{\Psi_{\mathrm{ref}}}$,
\begin{align}
    F(\hat{\rho},\Psi_{\mathrm{ref}}) \leq B_k(\Psi_{\mathrm{ref}}) = \sum_{i=1}^k \lambda_i^2,\label{eq:bounds}
\end{align}
where $k$ is the entanglement dimension, or Schmidt rank, of $\hat{\rho}$ and $\lambda_1^2\geq \lambda_2^2 \ldots\geq\lambda_N^2$ are the squared Schmidt coefficients of the reference state \cite{fickler_interface_2014}.
If Eq.~\eqref{eq:bounds} is violated for a given $k$, then the prepared state $\hat{\rho}$ is incompatible with less than $k+1$-dimensional entanglement.
Therefore, a fidelity measurement can be used directly to certify a lower bound on the entanglement dimension of the state.
For the reference-state family $\ket{\Psi_{\mathrm{ref}}(\boldsymbol{\lambda})}$ introduced above the fidelity is given by
\begin{align}
    F(\hat{\rho},\Psi_{\mathrm{ref}}) = \sum_{i,j=1}^N\lambda_i\lambda_j\bra{ii}\hat{\rho}\ket{jj}.\label{eq:fidelity}
\end{align}
However, fidelity measurements are costly and the experimental complexity scales unfavorably with the local Hilbert-space size \cite{bavaresco_measurements_2018}.

Here, we propose measuring a lower bound on the fidelity $F(\hat{\rho},\Psi_{\mathrm{ref}})$ in the spirit of~\cite{euler_detecting_2023}.
For the considered experimental platform, it is straightforward to take measurements in the time-bin basis $\bra{ij}\hat{\rho}\ket{ij}$ by recording the time bins $i$ and $j$ in which signal and idler photon arrive at the detector.
In addition, by subjecting the state to a DTQW before taking these measurements, it is also possible to obtain partial information on the coherences $\bra{ii}\hat{\rho}\ket{jj}$ appearing in the fidelity in Eq.\eqref{eq:fidelity}.
Due to the flexible programmability of the central coupler, a plethora of different interference patterns can be realized.
We can therefore choose to optimize the DTQW trajectories with respect to different key parameters, such as the required number of measurement settings, the bound tightness, or the necessary experimental shot budget.
Therefore, we propose two different measurement schemes which we outline in the following subsections, and we give a third more comprehensive scheme in Appendix~\ref{app:scheme}.

In the presence of time-bin-dependent initial-state phases, as in Eq.~\eqref{eq:phase-imprint}, an additional algorithmic step to learn and correct these phases has to be employed, significantly increasing the reference-state overlap.
We describe the details of this procedure in Sec.~\ref{sec:phase-correction}.

\subsection{Compound DTQW}
\label{sec:compound}

It is possible to design a set of comparatively simple DTQW trajectories using the internal structure of $\ket{\Psi_{\mathrm{ref}}}$.
Only coherences between states with signal and idler photon in the same time bin appear in $F(\hat{\rho},\Psi_{\mathrm{ref}})$, and hence only two different time bins contribute to any given coherence $\bra{ii}\hat{\rho}\ket{jj}$.
Consequently, all the necessary information can be extracted from two-bin interference patterns alone.
The simplest approach to measuring different combinations of two bins in parallel is to group all bins into pairs, which are then made to interfere pairwise before subsequent arrival-time readout.
To cover all $N(N-1)/2$ different combinations of two bins, it is sufficient to prepare $N-1$ different DTQW configurations (or $N$ for odd $N$, respectively), each achieving a different grouping of bins.
The necessary settings can be constructed for any $N$ by translating the task to an edge-coloring problem on the underlying fully connected graph of time bins \cite{edouard_recreations_1883, Baranyai_hypergraph_1975}.

\begin{figure}
    \centering
    \includegraphics[width=\linewidth]{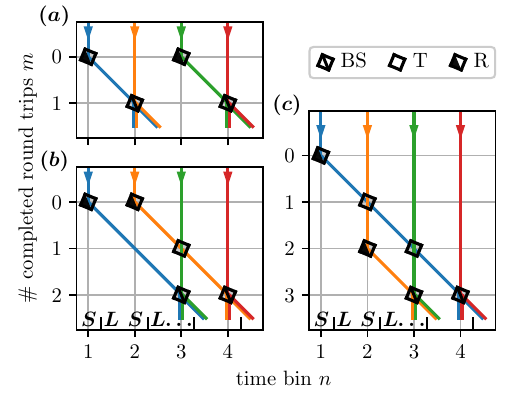}
    \caption{A diagrammatic illustration of the three settings of the compound DTQW scheme for $N=4$.
The different initial time-bin populations are represented by the four colored lines.
All bins start in the short loop (vertical lines).
Their trajectory in the DTQW can be traced along the vertical axis.
In all three cases, two earlier pulses are being injected into the long loop (diagonal lines) to delay them relative to the two other bins.
Once the two pairs coincide at the central coupler, they are made to interfere with a balanced beam-splitter setting and are subsequently ejected.
This scheme generates all combinations of two-bin pairings.
BS: balanced beam splitter, T: full transmission, R: full reflection }
    \label{fig:compound_all}
\end{figure}

A diagrammatic illustration of the three different settings needed for $N=4$ is given in Fig.~\ref{fig:compound_all}.
Each diagram describes the logical setting of the central coupler and the resulting time-bin trajectories as a function of time.
The four different time bins, represented by four different colors, are injected into the short loop at the top of each diagram as vertical lines.
In each setting, two bins are chosen and fully transferred to the long loop (depicted as diagonal lines) to remove the intrapair delay.
The timings are chosen such that all pairs interfere during the same round trip at the central coupler, which is then configured as a balanced beam splitter.
The number of round-trips in each setting (vertical depth of the diagram) scales with the maximum relative delay within each bin pair, as each round-trip in the long loop compensates a delay of $\Delta t$.

It is straightforward to compute the populations in the joint time-bin basis after each DTQW, as all pairs go through equivalent beam-splitter configurations.
Of special interest is the detection of both photons in the same DTQW branch in any of the diagrams.
Here, we label the probability of detecting both photons in the same output arm of the final beam splitter connecting bins $i$ and $j$ as $p_c(i,j)$.
Analogously, we refer to the probability for the detection of one photon in each arm of the same beam splitter as $p_{nc}(i,j)$.
The difference of these two probabilities $\Delta p_c(i,j)$ can be expressed as
\begin{align}
    \begin{split}
       \Delta p_c(i,j)\coloneqq& \,p_c(i,j) - p_{nc}(i,j)\\ =& 2\operatorname{Re}\left[\bra{ii}\hat{\rho}\ket{jj} + \bra{ij}\hat{\rho}\ket{ji}\right],\label{eq:compound}
     \end{split}
 \end{align}
with the full computation given in Appendix~\ref{app:DTQW}.
In comparisons, the reference state fidelity from Eq.~\eqref{eq:fidelity} for the case of real coefficients $\lambda_i\in\mathbb{R}$ can be expressed as
 \begin{align}
    F(\hat{\rho},\Psi_{\mathrm{ref}}) = \sum_i\lambda_i^2\bra{ii}\hat{\rho}\ket{ii} + 2\sum_{i < j}^N\lambda_i\lambda_j\operatorname{Re}(\bra{ii}\hat{\rho}\ket{jj}).\label{eq:fidelity_real} 
 \end{align}
 Therefore, only the first term appearing in $\Delta p_c(i,j)$, $\bra{ii}\hat{\rho}\ket{jj}$, contributes to the fidelity, whereas the other coherence, $\bra{ij}\hat{\rho}\ket{ji}$, does not.
We employ the strategy presented in \cite{euler_detecting_2023} to construct a fidelity bound by taking the sum over the differences of these two probabilities for all bin pairs and subsequently subtract the upper bound 
\begin{align}
    \begin{split}
        \operatorname{Re}\left[\bra{ij} \hat{\rho} \ket{ji} \right] &\leq |\bra{ij} \hat{\rho} \ket{ji}|\\
        &\leq \sqrt{\bra{ij} \hat{\rho} \ket{ij} \bra{ji} \hat{\rho} \ket{ji}}\\
        &\eqqcolon Z_{ij}^{ji}
    \end{split}\label{eq:Zij}
\end{align}
to remove the unwanted contribution of $\operatorname{Re}\left[\bra{ij} \hat{\rho} \ket{ji} \right]$.
This yields the final fidelity lower bound
\begin{align}
    \begin{split}
        \tilde{F}(\hat{\rho},\Psi_{\mathrm{ref}}) =& \sum_i\lambda_i^2\bra{ii}\hat{\rho}\ket{ii}\\
        + &\sum_{i<j}\lambda_i\lambda_j\left(\Delta p_c(i,j) - 2Z_{ij}^{ji}\right)\\
        \leq& F(\hat{\rho},\Psi_{\mathrm{ref}}) .\label{eq:bound_compound}
    \end{split}
\end{align}
Since every coherence $\bra{ii}\hat{\rho}\ket{jj}$ is measured and extracted independently, it is immediately possible to construct fidelity bounds to reference states of the form given in Eq.~\eqref{eq:reference_state} with individual real weights $\lambda_i$ for arbitrary $N$.
As mentioned before, it is always possible to obtain real weights from any generic state with, in general, complex weights by applying phase rotations to the individual basis states.
This enables treating reference states with arbitrary relative phases between the different time bins by preceding the DTQW with a dedicated phase-measurement step.
We cover this additional procedure, which can also be applied to other DTQW configurations, in Sec.~\ref{sec:phase-correction}.

Alternatively, instead of pairing each bin with only one other bin per DTQW, it is also conceivable to generate all pairings during the same DTQW setting.
This configuration allows for extracting information on even more coherences, but requires more roundtrips, and thus suffers from higher photon loss in practice.
A detailed discussion is provided in Appendix~\ref{app:scheme}.

\subsection{Single-Setting DTQW}
\label{sec:single-setting}

\begin{figure}
    \centering
    \includegraphics[width=\linewidth]{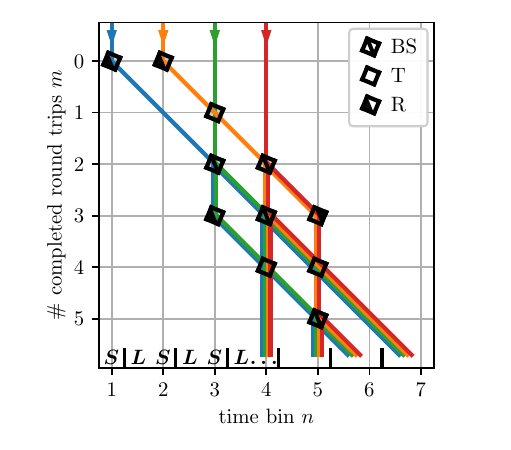}
    \caption{An illustration of the single-setup DTQW configuration for $N=4$.
The scheme interferes all bins in each of the $N$ output bins.}
    \label{fig:single_setup_N4}
\end{figure}

As an alternative, we also propose a single DTQW configuration that achieves full interference of all bins in all populated output bins; the corresponding diagram of the beam trajectories for $N=4$ is displayed in Fig.~\ref{fig:single_setup_N4}.
The configuration is built up recursively and can be constructed for cases with $N = 2^n,\,n\in\mathbb{N}$.
In this scheme, as the first step, the earlier half of the bins is delayed and interfered with the later half, generating $N$ outputs containing information of two time bins each.
Subsequently, we combine pairs of outputs, starting at the center in the $N$th bin, and then go outward symmetrically.
For $N=4$, a single additional beam-splitter pass for each pulse is sufficient to achieve full interference in every populated bin.
For larger $N$, the number of the interference events scales as $\sim \log_2(N)$, but in order to accommodate all trajectories, the total DTQW depth is $M=2(N-1)$.
We illustrate this further for the case of $N=8$ in Appendix~\ref{app:compound} in Fig.~\ref{fig:single_setup_N8}.

This DTQW configuration yields a total coincidence probability $p_c$, i.e., the probability to simultaneously detect the two photons in any of the output bins, which maps to the initial state as 
\begin{align}
    \begin{split}
        p_c &= \sum_{i=N}^{\mathclap{3N/2 - 1}} \bra{iSiS}\hat{U}\hat{\rho}\,\hat{U}^\dagger\ket{iSiS} + \sum_{\mathclap{i=3N/2}}^{2N - 1} \bra{iLiL}\hat{U}\hat{\rho}\,\hat{U}^\dagger\ket{iLiL}\\
        &= \frac1N\left(\sum_{i,j=1}^N\bra{ii}\hat{\rho}\ket{jj} \right) + \sum_{\substack{i\neq j\\ \lor k \neq l}}^N c_{ij}^{kl}\bra{ij}\hat{\rho}\ket{kl}\\
        &= F(\hat{\rho},\Psi_{\mathrm{MES}}) + \sum_{\substack{i\neq j\\ \lor k \neq l}}^N c_{ij}^{kl}\bra{ij}\hat{\rho}\ket{kl}
    \end{split}
\end{align}
with coefficients $c_{ij}^{kl}$ depending on $N$.
Here, $\ket{iSiS}$ and $\ket{iLiL}$ mark the state of both photons in the $i$th bin and both propagating in the short or long loop, respectively.
Importantly, the first sum containing the terms contributing to the fidelity always has a prefactor $1/N$ guaranteed by the symmetry of the DTQW.
This holds true as the number of loop switches along the DTQW trajectory of a bin contributing to a pulse in the short loop is always even and always odd for a pulse in the long loop.
Thus, the difference in the number of loop switches between bins $i$ and $j$ is even in each pulse, that is, $n_i - n_j = 2c,~c\in\mathbb{Z}$.
With each switch, each photon picks up a phase of $\pi/2$.
Taking into account the form of the two-photon coherences, $\bra{ii}\hat{\rho}\ket{jj}$, the corresponding phase factor is thus always $\exp(i2(n_i - n_j) \frac{\pi}{2}) = \exp(i2\pi c) = 1$, which ensures that all relevant coherences have the same real and positive weight.

Therefore, the first sum reassembles the fidelity $F(\hat{\rho},\Psi_{\mathrm{MES}})$ to the maximally entangled state (MES) $\ket{\Psi_{\mathrm{MES}}}=1/N\sum_{i=1}^N\ket{ii}$.
Thus, we can extract the fidelity from a single setting by utilizing the bound as described in the preceding section, yielding the simplified bound
\begin{align}
    F(\hat{\rho},\Psi_{\mathrm{MES}}) \geq  \tilde{F}(\hat{\rho},\Psi_{\mathrm{MES}}) = p_c - \sum_{\substack{i\neq j\\ \lor k \neq l}}^N \bigl|c_{ij}^{kl}\bigr| Z_{ij}^{kl}. \label{eq:bound_single}
\end{align}
This simplifies the measurement protocol to a single DTQW setting independent of $N$, while the compound approach employs $N-1$ or $N$ settings, respectively, in addition to the direct population measurements needed to determine the $Z_{ij}^{kl}$ in both cases.
Consequently, all experimental samples dedicated to measuring the DTQW outcomes can be spent in a single setting, lowering overall statistical requirements.
This simplicity comes at the cost of reduced flexibility, as only $\ket{\Psi_{\mathrm{MES}}}$, i.e., a flat distribution of Schmidt coefficients (but potentially arbitrary phases), can be used as the reference.
However, previous studies found that the use of $\ket{\Psi_{\mathrm{MES}}}$ in this capacity yields favorable bounds $B_k$ for $k\approx N$, such that states with a high entanglement dimension can be detected more robustly \cite{bavaresco_measurements_2018, euler_detecting_2023}.
Thus, giving up flexibility in favor of simplicity might be appropriate in some circumstances.

\subsection{Phase-correction protocol}
\label{sec:phase-correction}

For general initial quantum states $\hat{\rho}$, it is insufficient to consider reference states with real coefficients $\lambda_i$.
Deviations between the phases of the prepared state $\rho$, and the ones of the reference state lead to rapidly diminishing state overlap, and thus to the loss of any detectable entanglement.
By taking the phase-sensitive coherences contributing to the fidelity and expanding them for general coefficients,
\begin{align}
    \begin{split}
        \sum_{i\neq j}^N&\lambda_i^{*}\lambda_j\bra{ii}\hat{\rho}\ket{jj}=\\
                    2\sum_{i<j}^N&\Big[\operatorname{Re}(\lambda_i)\operatorname{Re}(\lambda_j) + \operatorname{Im}(\lambda_i)\operatorname{Im}(\lambda_j)\Big]\operatorname{Re}\!\left(\bra{ii}\hat{\rho}\ket{jj}\right)\\
        +  &\Big[\operatorname{Im}(\lambda_i)\operatorname{Re}(\lambda_j) - \operatorname{Re}(\lambda_i)\operatorname{Im}(\lambda_j)\Big]\operatorname{Im}\!\left(\bra{ii}\hat{\rho}\ket{jj}\right),\label{eq:fidelity_complex}
        \end{split}
\end{align}
one can see that the imaginary parts of the coherences also contribute to the fidelity in the case of non-zero relative phases.
The first straightforward approach to overcome this problem is simply to find bounds for the imaginary parts as well.
This can be realized for the compound DTQW scheme by repeating each setting two more times with different initial-state phase imprints.
In the first iteration, a phase of $\varphi_j=\pi/2$ is applied to the later bin of all pairs in each setting, while in the second iteration, $\varphi_j=\pi/4$ is imprinted.
The resulting probability differences between coincidence- and no-coincidence detections for bins $i$ and $j$ are
\begin{subequations}
    \begin{align}
        \begin{split}
             \Delta p_c(i,j)_{\big|\varphi_j = \frac \pi2} &= 2\operatorname{Re}\Bigl[e^{-i\pi}\bra{ii}\hat{\rho}\ket{jj} + \bra{ij}\hat{\rho}\ket{ji}\Bigr]\\
            &= 2\left( - \operatorname{Re}\left[\bra{ii}\hat{\rho}\ket{jj}\right] + \operatorname{Re}\left[\bra{ij}\hat{\rho}\ket{ji}\right]\right),
        \end{split}\\
        \begin{split}
             \Delta p_c(i,j)_{\big|\varphi_j = \frac \pi4} &= 2\operatorname{Re}\left[e^{-i\pi/2}\bra{ii}\hat{\rho}\ket{jj} + \bra{ij}\hat{\rho}\ket{ji}\right]\\
             &= 2\left(\operatorname{Im}\left[\bra{ii}\hat{\rho}\ket{jj}\right] + \operatorname{Re}\left[\bra{ij}\hat{\rho}\ket{ji}\right]\right),
         \end{split}
         \label{eq:compound_phase}
    \end{align}
\end{subequations}
where all phases cancel out on the second coherence.
Combining these measurements with the phase-free result from Eq.~\eqref{eq:compound}, one obtains
\begin{subequations}
    \begin{align}
         \operatorname{Re}\left[\bra{ii}\hat{\rho}\ket{jj}\right] =&\frac{\Delta p_c(i,j) -  \Delta p_c(i,j)_{\big|\varphi_j = \frac \pi2}}{4},\\
         \begin{split}
             \operatorname{Im}\left[\bra{ii}\hat{\rho}\ket{jj}\right] =& - \frac{\Delta p_c(i,j) + \Delta p_c(i,j)_{\big|\varphi_j = \frac \pi2}}{4}\\
             &+\frac{\Delta p_c(i,j)_{\big|\varphi_j = \frac \pi4}}{2}.
         \end{split}
    \end{align}
    \label{eq:real_imag}
\end{subequations}
Thus, it is possible to not only bound but also precisely measure the real and imaginary parts of all relevant coherences.
However, the associated experimental complexity in terms of required measurement repetitions increases threefold, and the method is only applicable to the compound DTQW scheme introduced in Sec.~\ref{sec:compound}.
This makes this approach, though powerful, challenging to implement, especially for large $N$.

\begin{figure}
    \centering
    \includegraphics[width=\linewidth]{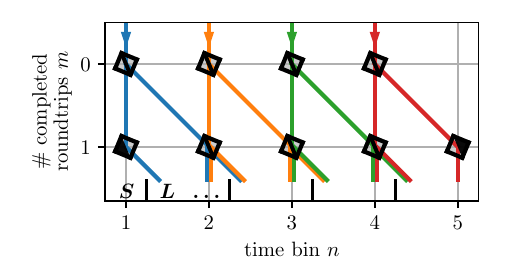}
    \caption{An illustration of the DTQW configuration used to estimate the relative phases between adjacent time bins.}
    \label{fig:phase_N4}
\end{figure}

Because of these limitations, we propose an alternative method to estimate the relative phases of the photon-pair state, which is based on the assumption that the phase of each two-photon state can be parametrized as $\varphi_{ij}=\varphi_i+\varphi_j$, i.e., phase imprints are assumed to be uncorrelated between signal and idler photon.
Then, instead of measuring the real and imaginary parts of every contributing coherence, an initial phase-estimation protocol is applied that just needs to determine relative phases between adjacent time bins.
This approach requires only two round trips independent of $N$.
We utilize a single DTQW configuration that generates all nearest-neighbor pairings; an illustration for $N=4$ is shown in Fig.~\ref{fig:phase_N4}.
The configuration is run twice; first with no phase imprints, then with a phase gradient that implements $\pi/4$ phase increments between adjacent time bins.
Thus, in these two measurement settings one can measure adjacent bin detection probability differences $\Delta p_c(i,i+1)$ and $\Delta p_c(i,i+1)_{\big|\varphi_{i+1} = \frac \pi4}$ as given in Eqs.~\eqref{eq:compound} and \eqref{eq:compound_phase}.
Utilizing the bound introduced in Eq.\eqref{eq:Zij}, we can use these rates to lower bound the respective real and imaginary parts of the contributing coherences,
\begin{subequations}
    \begin{align}
        \operatorname{Re}\left[\bra{ii}\hat{\rho}\ket{i'i'}\right] &\geq \left(\frac{\Delta p_c(i,i')}{2} - Z_{ii'}^{i'i}\right)\eqqcolon c_R,\\
        \operatorname{Im}\left[\bra{ii}\hat{\rho}\ket{i'i'}\right] &\geq \left(\frac{\Delta p_c(i,i')_{\big|\varphi_{i'} = \frac \pi4}}{2} -  Z_{ii'}^{i'i}\right)\eqqcolon c_I,
    \end{align}
\end{subequations}
with $i'=i+1$ for brevity of notation.
Therefore, for $|\!\bra{ii}\hat{\rho}\ket{i'i'}\!| \gg |\!\bra{ii'}\hat{\rho}\ket{i'i}\!|$, we can approximate the phase of all $\bra{ii}\hat{\rho}\ket{i'i'}$ as
\begin{equation}
    \phi_{i,i'} \approx \operatorname{atan2}\left(c_I, c_R\right).\label{eq:phase_approx}
\end{equation}
By setting $\varphi_j = -\sum_{i<j} \phi_{i,i'}/2$ as the initial-state phase in the following, we can effectively rotate the relevant coherences back into the real plane.
As this operation does not change the entanglement spectrum, we can still perform the certification of the entanglement dimension on the rotated state.
This can significantly increase the fidelity between $\hat{\rho}$ and reference states with purely real coefficients $\lambda_i$ in the case of phase mismatch, enhancing the capability to detect high-dimensional entanglement for both measurement schemes discussed in Secs.~\ref{sec:compound} and~\ref{sec:single-setting}.
Finally, it should be stressed that if the phase parametrization or Eq.~\eqref{eq:phase_approx} do not hold, it is possible that a deviating phase profile is learned and applied to the state.
However, the certified fidelity and thus detected entanglement dimension are always rigorous lower bounds on the underlying true values, as the applied rotations fall under the LOCC paradigm \cite{nielsen_conditions_1999}.

\section{Numerical simulations}
\label{sec:numerics}

To benchmark the performance of the proposed measurement schemes under experimentally realistic conditions, we conducted a suite of numerical simulations.
We consider three main error sources, which are photon loss in combination with dark counts, statistical errors, and central-coupler configuration noise.
We first go into more detail on the error model assumptions, and then present numerical results for the different detection schemes.

The dominant error channel of the considered platform is photon loss in one of the optical couplers or in the optical fiber.
For the concrete experiment reported in Ref.~\cite{monika_quantum_2024}, the authors state a loss of $1.84\,\mathrm{dB}$ per photon round trip plus an additional loss of $0.7\,\mathrm{dB}$ for the out-coupling of the state after the DTQW.
The main effect is an increase in the amount of measurement statistics needed, since events with a number of detected photons other than two, i.e., photon loss or multi-pair generation, can be filtered out in postselection and do not impair the correct detection of high-dimensional entanglement.
Only single-photon loss in combination with a detector dark count is indistinguishable from a regular two-photon event and can be described as an effective dephasing channel applied to the target state $\ket{\Psi}$, producing $\hat{\rho} = (1-\alpha)\ket{\Psi}\!\bra{\Psi} + \frac{\alpha}{N^2} \mathbb{1}$.
Here, we consider both pure states and dephased states, where for a given $N$, $\alpha$ is chosen such that the state has a purity of $\operatorname{tr}(\hat{\rho}^2)\approx0.9$.

Finally, we assume random noise on the central optical switch.
We parameterize the coupler through a switching angle, with $\theta\in[0,\pi/2]$ (for more details on this, see Appendix~\ref{app:DTQW}).
In all data shown, we assume that each angle $\tilde{\theta}$ realized experimentally is subject to random noise and therefore sampled uniformly according to $\tilde{\theta}\sim\mathcal{U}[\theta-0.02\pi,\theta+0.02\pi]$.

\subsection{Sampling statistics}
\label{sec:sampling}

\begin{figure}
    \centering
    \includegraphics[width=\linewidth]{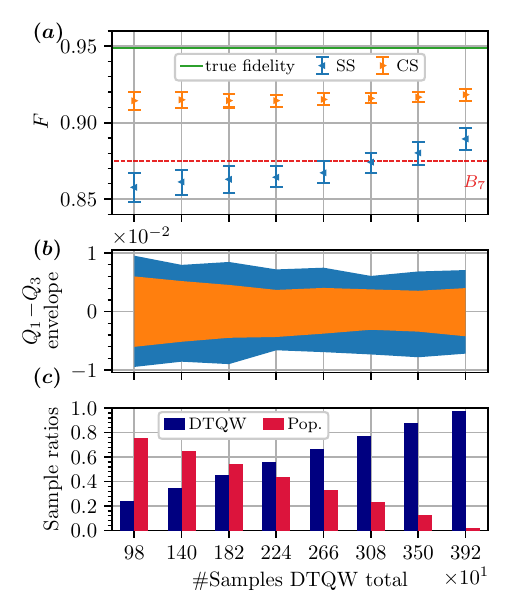}
    \caption{The fidelity-bound performance for different distributions of the measurement budget on a phase-free dephased initial state.
    (a) Certified fidelity-bound medians (SS: Single setting, CS: Compound setting) and true fidelity to $\ket{\Psi_{\mathrm{MES}}}$ for different distributions of the experimental samples.
    The errorbars correspond to the interquartile range $Q_1 - Q_3$.
    (b) Corresponding $25\%-75\%$ confidence envelopes relative to the medians from (a).
    (c) Proportions of number of samples used for DTQW and population measurements.
    All data has been acquired for $N=8$ with a fixed $N_{\mathrm{sample}}=4000$.}
    \label{fig:FidelityEstimate}
\end{figure}

Here, we investigate how limited measurement statistics impact the certification ability of the proposed method.
In a first step, we simulate the application of both the compound and single-setting implementations to a dephased maximally entangled state with $N=8$ and no relative phases between the different time bins.
To determine the sensitivity of the protocol to finite statistics, we fixed the total number of experimental state realizations to $N_{\mathrm{sample}}=4000$. This total shot budget can be distributed between measurements of the time-bin populations (without DTQW stage) and coherence measurements involving one of the DTQW schemes described in Sec.~\ref{sec:results}.
The resulting certified fidelities and entanglement dimensions for different experimental budget splits are shown in Fig.~\ref{fig:FidelityEstimate}.

We observe only weak dependence of the fidelity bounds generated through the compound method on the actual distribution of the samples, even in the limit of very low sample numbers for population measurements, as seen in Fig.~\ref{fig:FidelityEstimate}(a).
However, for the single-setting scheme, we find a bias towards higher certified fidelities in this limit, which is caused by insufficient resolution to resolve the small state populations related to photon loss and subsequent detector dark count.
Thus, these populations and the derived correction terms $Z_{ij}^{kl}$ [cf. Eq.~\eqref{eq:Zij}] are being underestimated, which leads to an erroneous overestimation of the certified fidelity.
The compound scheme is less susceptible to low population sample numbers, as fewer terms remain in the sum of coincidence rates that must be considered. State dephasing and the simulated noise on the central coupler lead to the gap between the true fidelity and the fidelity bounds.

The corresponding quartiles $Q_1$ and $Q_3$ relative to the median are plotted in Fig.~\ref{fig:FidelityEstimate}(b).
Across all distribution ratios, the compound method outperforms the single-setting method, and we observe a trend for both towards smaller errors for higher DTQW-sample numbers.

Based on these findings, for all the following simulations with $N=8$, we dedicate $N_{\mathrm{sample}}^{\mathrm{pop}}=2000$ to population measurements to mitigate undersampling while keeping statistical errors manageable.

\begin{figure}
    \centering
    \includegraphics[width=\linewidth]{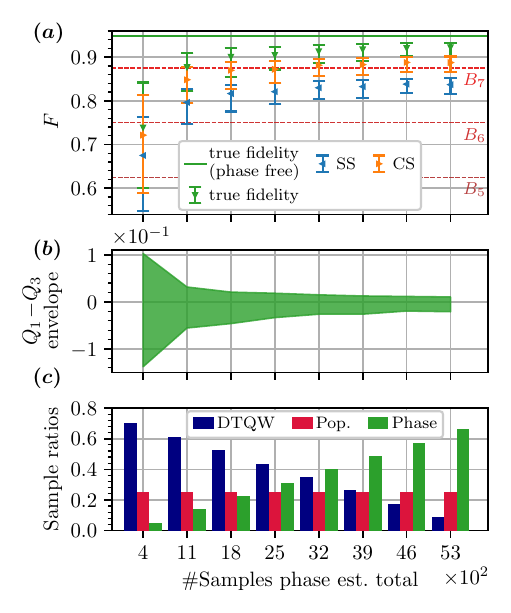}
    \caption{The fidelity-bound performance for different distributions of the measurement budget on a random-phase dephased state.
    (a) Certified fidelity-bound medians (SS: Single setting, CS: Compound setting) and true fidelity of the phase-corrected and phase-free prepared state to $\ket{\Psi_{\mathrm{MES}}}$ as a function of experimental budget split.
    The errorbars correspond to the interquartile range $Q_1 - Q_3$.
    (b) Corresponding $25\%-75\%$ confidence envelope of the true fidelity of the phase-corrected state relative to the median from (a). The envelopes for the compound- and single-setup schemes have visually indiscernible quartile envelopes and are omitted here for clarity. 
    (c) Proportions of number of samples used for DTQW, phase, and population measurements.
    All data has been acquired for $N=8$ with a fixed $N_{\mathrm{sample}}=8000$, $25\%$ of which are reserved for population measurements.}
    \label{fig:PhaseEstimate}
\end{figure}

In a second step, we introduce random phases between different time bins in the initial state, necessitating phase-correction before any DTQW readout.
Therefore, we consider an increased  number of samples of $N_{\mathrm{sample}}=8000$ of which $6000$ are distributed among phase- and DTQW measurements; the certified fidelities and corresponding sample distributions for a number of different sample allocations are shown in Fig.~\ref{fig:PhaseEstimate}.
We observe a significant increase in the true fidelity between the phase-corrected prepared state and the phase-free maximally entangled state by increasing the number of phase-correction samples.
Furthermore, the observed error of the true phase-corrected state fidelity and of the fidelity bounds are up to one order of magnitude higher compared to the phase-free case [cf. Figs.~\ref{fig:FidelityEstimate}(b) and~\ref{fig:PhaseEstimate}(b)], meaning that the phase correction procedure is the dominant source of uncertainty for both fidelity bounds.
This suggests utilizing a substantial amount of samples for phase estimation and correction to both increase certified fidelities bounds and reduce the corresponding errors.

In the following, we dedicate half of all available samples for phase estimation and divide the remaining samples evenly between the time-bin populations and the DTQW measurements.

\subsection{Scaling with number of time bins \textit{N}}

\begin{figure}
    \centering
    \includegraphics[width=\linewidth]{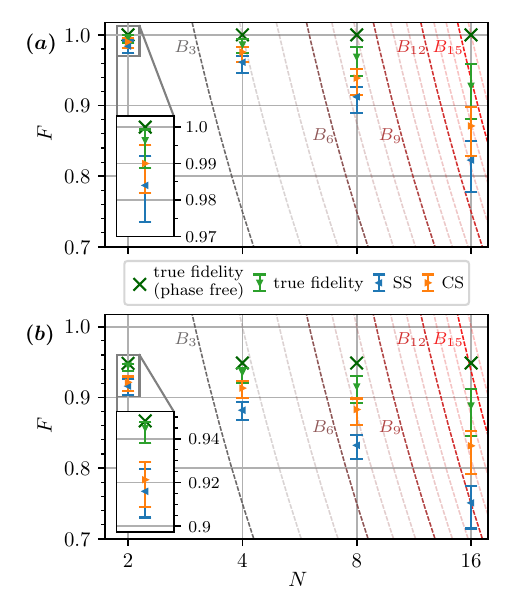}
    \caption{The fidelity-bound performance with phase-correction for different values of $N$ (SS: Single setting, CS: Compound setting).
    (a) Certified fidelities on a pure maximally entangled state with random phases.
    (b) Certified fidelities on a dephased maximally entangled state with random phases.
    The dashed lines are the $B_k$ entanglement-dimension thresholds.
    All data has been acquired with a fixed $N_{\mathrm{sample}}=8000$, with a 1:1:2 split among population, DTQW, and phase measurements.}
    \label{fig:N_scaling}
\end{figure}

We simulate the application of the method including phase correction for $N\in\{2,4,8,16\}$ for both pure and dephased states to investigate how the bound performance scales with $N$.
To account for the increasing sample requirements with larger $N$, we scale the number of samples linearly with $N$, i.e., \hbox{$N_\mathrm{samples} = 1000N$}.
The certified fidelities are given in Fig.~\ref{fig:N_scaling}.

We observe qualitatively similar behavior between the results for pure states in Fig.~\ref{fig:N_scaling}(a) and the dephased states in Fig.~\ref{fig:N_scaling}(b), with certified fidelities decreasing with larger $N$ in both cases.
For $N\in\{2,4\}$, statistical errors on the compound- and single-setting bound dominate, while for larger $N$ the phase correction is the leading cause of uncertainty, as previously discussed.
The dephasing introduces a nearly constant offset in the phase-corrected true fidelity across all $N$, which then propagates to the certified fidelities.

Both bounds certify the full entanglement dimension with at least 75 \% confidence for $N\in\{2,4,8\}$ in the pure case and for $N\in\{2,4\}$ in the dephased case, with 7-dimensional entanglement certified for $N=8$.
For $N=16$, we see that the compound method outperforms the single setting method by one entanglement dimension in both scenarios.
Here, the observed drop in the phase-corrected true fidelities indicates that increases in the number of samples could cause further improvements in both the fidelity-bound median and uncertainty, respectively.

\section{Entanglement detection for multi-photon-pair states}
\label{sec:multi-spdc}

Up until now, we have considered the standard scenario where a single photon pair is produced through SPDC.
However, at increased laser power, higher-order contributions may become significant.
It is conceivable, albeit technically challenging, to postselect measurement results where two photon pairs are detected.
The corresponding multi-photon quantum state
\begin{align}
    \ket{\Phi(\boldsymbol{\alpha})}_{\mathrm{ref}}=\sum_{j\leq k}^N \alpha_{jk} \ket{jk}_S\otimes\ket{jk}_I\eqqcolon\sum_{j\leq k}^N \alpha_{jk} \ket{jkjk}\label{eq:state-multi-spdc}
\end{align}
with two indistinguishable signal and idler photons has an increased entanglement dimension of $D_{\mathrm{ent}}=\binom{N + 1}{2}=N(N + 1)/2$ between the two different photon species.
To emphasize the generality of our method, we illustrate how the methods presented in this work can be adapted to also allow the detection of high-dimensional bipartite entanglement for multi-photon states.

\begin{figure}
    \centering
    \includegraphics[width=\linewidth]{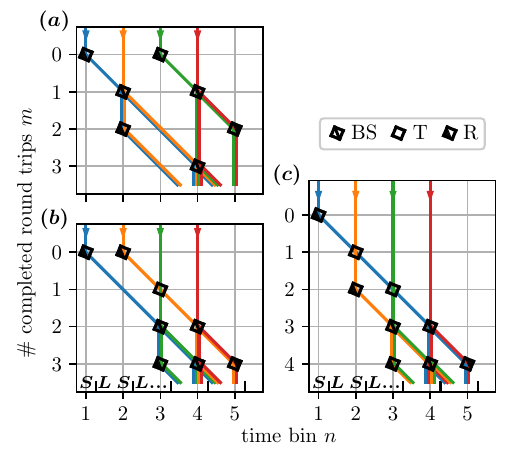}
    \caption{A diagrammatic illustration of the three additional DTQW schemes for two photon pairs for $N=4$.
    This scheme generates four-bin interference in each diagram.
    The diagrams can be obtained from those in Fig.~\ref{fig:compound_all} by adding one final balanced beam splitter in the central bin.}
    \label{fig:compound-multi-spdc}
\end{figure}

The fidelity between the above reference state and the postselected four-photon state is given by
\begin{align}
     F(\hat{\rho},\Phi_{\mathrm{ref}}) = \sum_{i\leq j}^N \sum_{k\leq l}^N\lambda_{ij} \lambda_{kl} \bra{ijij} \hat{\rho} \ket{klkl},\label{eq:fidelity-multi-spdc}
\end{align}
where up to four different time bins contribute to a single coherence.
It is therefore necessary to create four-bin interference to capture the full quantum correlations in this family of states.
We propose adapting the compound method introduced in Sec.~\ref{sec:compound} for four bins, such that in each setting the two bin pairs are interfered at the end to achieve four-bin interference in the central bin.
The three derived settings are displayed in Fig.~\ref{fig:compound-multi-spdc}.
In each of the three diagrams, the time bins are subjected to a different DTQW trajectory, leading to different relative phases in the final beam output.
This can be utilized to cancel out certain non-contributing coherences by combining coincidence-rate measurements from all three configurations.

For $N>4$, this pattern would have to be measured for all combinations of four bins.
Here, we want to focus on the case of $N=4$, such that the three DTQW settings shown already include all the necessary information for a fidelity bound.
In the following, three detection patterns are relevant.
First, all four photons are detected in the same output of the final beam splitter.
We label the probability of such a full coincidence (FC) event as $p_{FC}^{\alpha}$, where $\alpha\in\{a,b,c\}$ corresponds to the three diagrams in Fig.~\ref{fig:compound-multi-spdc}.
Second, there can be species coincidence (SC), that is, the two signal photons are detected in one output and the two idler photons in the other, with a corresponding probability $p_{SC}^{\alpha}$.
Finally, one can detect signal-idler-pair coincidence (PC), with a signal-idler pair detected in each beam output with probability $p_{PC}^{\alpha}$.
As we show in Appendix~\ref{app:multi-spdc}, one can combine the measurement of these event rates with relevant four-photon populations and measurements taken in the compound setup introduced in Sec.~\ref{sec:compound}.
There, we also measure the probability of detecting a signal-idler pair in each output arm after path interference, analogously labeled $p_{PC}(i,j)$ for photons originating from bins $i$ and $j$.
Thus, one obtains the compound quantity
    \begin{align}
        \begin{split}
           \mathcal{C}=&8\sum_\alpha p_{FC}^{\alpha} - p_{SC}^{\alpha} + 2p_{PC}^{\alpha} + \sum_{i<j}^N p_{PC}(i,j)\\
           &-\frac12 \sum_i\bra{iiii} \hat{\rho} \ket{iiii} \\
            =&\sum_{\substack{i\leq j\\ k\leq l}}^N \bra{ijij} \hat{\rho} \ket{klkl} +\sum_{\substack{i\leq j\\ k\leq l}}^N \sum_{\substack{m\leq n\\ o\leq p}}^N c_{ijmn}^{klop} \bra{ijmn} \hat{\rho} \ket{klop}\label{eq:multi-spdc-C},
        \end{split}
    \end{align}
with matrix-element weights $c_{ijmn}^{klop}$ and the second sum on the RHS omitting all matrix elements already included in the first sum.
Importantly, the first sum is proportional to the fidelity to the maximally entangled four-photon state [see Eq.~\eqref{eq:fidelity-multi-spdc}].
By employing the same techniques as described above for the two-photon case, we can generate an analogous lower bound on this fidelity from the combined measurement result, which reads
\begin{align}
    \tilde{F}(\hat{\rho},\Phi_{\mathrm{MES}})= \frac{2}{N(N+1)}\left( \mathcal{C} - \sum_{\substack{i\leq j\\ k\leq l}}^N \sum_{\substack{m\leq n\\ o\leq p}}^N \Bigl|c_{ijmn}^{klop}\Bigr| Z_{ijmn}^{klop}\right),
\end{align}
where $Z_{ijmn}^{klop}$ has been defined in analogy to Eq.~\eqref{eq:Zij}.

To estimate the achievable fidelity through this bound, we apply it to a dephased maximally entangled four-photon state of $N=4$ and a purity of $\operatorname{Tr}(\hat{\rho}^2)\approx0.9$, but neglect finite measurement statistics and noise on the central coupler.
In this configuration, we find a lower fidelity bound of $\tilde{F}(\hat{\rho},\Phi_{\mathrm{MES}})\approx0.609$, which corresponds to a certifiable entanglement dimension of $D_{\mathrm{ent}}=7$.

One way to improve upon this result is to run each of the three four-bin DTQWs in four different phase patterns instead of a single, phase-free configuration.
We propose imprinting a phase of $\varphi_i=\pi$ on a different bin in each run and subsequently taking the sum over the four runs.
The relative phases cause the cancellation of terms of the form $\bra{ijij}\hat{\rho}\ket{lmln}$, i.e., terms with two uncorrelated photons in bins $n$ and $m$ in the bra or ket.
These are related to events with a single produced SPDC pair and two dark counts or two SPDC pairs with  single photon loss and one dark count in different time bins.
They are the dominant error source, as the corresponding bound is large, i.e., $Z_{ijij}^{lmln}\propto\sqrt{\bra{ijij}\hat{\rho}\ket{ijij}}$.
Applying this modified bound to the same state as above yields an increased $\tilde{F}(\hat{\rho},\Phi_{\mathrm{MES}})\approx0.915$ and a maximum certifiable $D_{\mathrm{ent}}=10$.
For more details, we again refer to Appendix~\ref{app:multi-spdc}.

These results demonstrate that the entanglement-certification method can in principle be extended to multi-photon scenarios as well, fully utilizing the flexibility of the platform.
However, we want to stress that a practical near-term implementation is facing significant technical challenges.
First, the production of two SPDC pairs instead of one from one pulse sequence is strongly suppressed, substantially increasing the necessary measurement times.
Furthermore, as the survival probability of the full quantum state decreases exponentially with photon number, feasibly attainable noise levels make the implementation of a four-photon certification scheme exceedingly difficult in state-of-the-art experiments.
However, as mentioned above, other discrete-quantum-walk platforms like bosonic atoms in optical lattices might have better noise characteristics such that near-term implementations in other systems might be more viable.
\section{Conclusions}
\label{sec:discussion}
In this work we showed that optical-fiber-loop platforms can be used to flexibly prepare and certify high-dimensional entanglement on the same device.
The certification procedures can be adapted to optimize different experimental aspects, such as the number of necessary measurement settings, the tightness of the bound, or the extent of extracted information.
By combining our entanglement-detection method with a phase-correction protocol, high-dimensional entanglement can be certified for initial states with arbitrary phase profiles.
In numerical simulations, the method has demonstrated robustness under constrained experimental shot budgets and representative experimental noise models.

The platform in combination with the detection method offers some key experimental advantages.
As the same fiber loops are used for state preparation and path interference, no major further measures for interferometer stabilization have to be taken.
Two single-photon detectors and two unbalanced fiber loops are sufficient to detect high-dimensional entanglement for any time-bin number $N$, making the platform compact, broadly configurable, and versatile in terms of the class of preparable and detectable states.

By reprogramming the central coupler, one can gain further information about a large number of time-bin coherences, including their real- and imaginary parts.
However, full quantum state tomography is not achievable with the proposed experimental capabilities.
Here, one would need the ability to imprint phases on the two photons independently \cite{takesue_implementation_2009}, which could be realized through potential differences in polarization after SPDC.

The method presented in this work belongs to a family of similar techniques that rely on complementary measurements in two or more bases \cite{sanpera_schmidt-number_2001, sentis_quantifying_2016, erker_quantifying_2017, bavaresco_measurements_2018, euler_detecting_2023}.
These techniques keep the experimental complexity and resource requirements low by capitalizing on the large degree of prior knowledge about the prepared state.
Another approach to high-dimensional entanglement certification is to measure correlations after random unitary application, which has recently been performed experimentally for the first time \cite{liu_characterizing_2023, wyderka_probing_2023, lib_experimental_2024}.
These methods do not require a common reference between parties; however, as we certify on the same device, this is not an issue here.
Additionally, our simulations show a very competitive bound performance using similar measurement statistics as Ref.~\cite{lib_experimental_2024}.

One of the major open challenges of our approach is photon loss, caused predominantly by the repeated fiber-coupler passes in the loops.
Crucially, the projected photon loss of the three stages of the protocol scales quite differently with $N$, as the loss strongly correlates with the necessary DTQW depth.
Phase measurements always require $M=2$ round trips independent of $N$, one for loading the state into the fiber loops and one for path interference.
Moreover, population measurements bypass the loops and thus have comparatively low loss rates.
Thus, both sampling processes are subject to constant loss for all $N$, and higher sample numbers for these parts can be achieved relatively easily, even for large $N$.
However, the round-trip depth of the proposed DTQW measurements scales linearly with $N$, leading to an exponentially decaying signal-to-noise ratio.
At the reported loss rates of $1.84\,$dB per photon round trip and a constant coupling loss of $0.7\,$dB, this would mean a worst-case loss of $8.06\,$dB for one of the $N-1$ settings of the compound method for $N=4$, corresponding to a two-photon survival probability of $\approx2.44\%$.
However, for $N=16$, this increases dramatically to $30.14\,$dB and an associated two-photon survival probability of $\approx10^{-4}\%$.
As a consequence, sample acquisition times increase exponentially, making the implementation of $N=16$ significantly more challenging at the same noise level.
However, DTQW depths of $M=4$ have already been demonstrated in the current hardware configuration.

To enable the preparation and certification of quantum states with even higher entanglement dimension,  a number of straightforward technical improvements may be implemented.
First, the current fiber-loop setup only supports $N=4$ time bins simultaneously.
This limitation can be lifted by increasing the length of both loops by the same multiple of the current length difference $\Delta L$.
As fiber losses are of the order of $\approx0.2\,$dB/km, the corresponding increase in loss would not be prohibitive compared to the current round-trip loss.
Here, technological advances are essential to reduce the fiber-coupler insertion loss to $\ll 1\,$dB to make data-acquisition times more practical.
Further performance gains might be achieved by reducing the response time of the central coupler of currently $\approx60\,$ns, which necessitates the long time-bin spacing $\Delta t=100\,$ns and brings the total time for state preparation and DTQW to the order of $\mathcal{O}(1\,$µs).
Thus, faster switching times would allow for the implementation of a much lower $\Delta t$ and generally faster state preparation and data collection.
These prospects show a clear path forward to use fiber-loop-generated high-dimensional entanglement for quantum technology applications.

\section*{Acknowledgments}

We thank M. C. Ponce, V. R. Kaipalath, L. J. Gonzalez Martin Del Campo, P. Emonts, and F. Steinlechner for fruitful discussions.
This research is supported by funding from the German Research Foundation (DFG) under the project identifier 398816777-SFB 1375 (NOA).

\appendix
\numberwithin{equation}{section}

\section{Model of DTQWs}
\label{app:DTQW}
Here, we briefly summarize the underlying mathematical model of the DTQWs considered in this work.
The following has been described in previous works, e.g., in Ref.~\cite{monika_quantum_2024}, but we give it here for completeness.

To fully describe the evolution of the state during the quantum walk, each pulse gets an additional label $S$ or $L$ to indicate whether it propagates in the short loop or in the long loop.
Thus, immediately after the quantum state has been injected into the short loop, the two photons are in the state
\begin{align}
    \ket{\Psi(\boldsymbol{\alpha})}=\sum_{j=1}^N \alpha_j\ket{j,S}_S\otimes\ket{j,S}_I,
\end{align}
where we have also reintroduced the signal- and idler-photon labels temporarily.
During a DTQW, two types of operators are applied to the quantum state in an alternating fashion: the coin operator $\hat{C}_m$ and the temporal shift operator $\hat{S}_m$, with $m$ being the current round-trip index.
The coin operator represents the central beam coupler and encodes how the incoming pulses of light are split between the long and short loops.
As the configuration of the central coupler can be changed between adjacent time bins, the corresponding coin operator $\hat{C}_m$ is of the form $\hat{C}_m = \sum_{n=1}^{N_m}\hat{C}_{m, n}$, with $\hat{C}_{m, n}$ being the coin operator for the $n$th bin out of $N_m$ on the $m$th round trip.
The symmetric single-photon coin operator realized by the central coupler used in Ref.~\cite{monika_quantum_2024} is given by
\begin{gather}
    \begin{aligned}
        \hat{C}_{m,n}^{sp} = \cos(\theta_{m,n})(\ket{n,S}\!\bra{n,S} &+ \ket{n,L}\!\bra{n,L})\\
        + i \sin(\theta_{m,n})(\ket{n,S}\!\bra{n,L} &+ \ket{n,L}\!\bra{n,S}),\label{eq:coin}
    \end{aligned}
\end{gather}
where $\theta_{m,n}\in[0,\pi/2]$ parametrizes the behavior of the coupler with full reflection, that is, each pulse stays in the current loop, at $\theta_{m,n}=0$ to full transmission, i.e., each pulse is completely transferred to the other loop, at $\theta_{m,n}=\pi/2$.
The full two-photon operator is then simply constructed as the product operator $\hat{C}_{m, n} = \hat{C}_{m,n,S}^{sp}\otimes\hat{C}_{m,n,I}^{sp}$.

The shift operator $\hat{S}_m$ implements the delay caused by the length imbalance of the two loops.
For the $m$th round-trip, the single-photon operator reads
\begin{align}
    \hat{S}_m^{sp} = \sum_{n=1}^{N_m}\ket{n,S}\!\bra{n,S} + \ket{n+1,L}\!\bra{n,L},\label{eq:shift}
\end{align}
only shifting pulses in the long loop by one bin.
Again, the full two-photon operator is the tensor product of the two single-photon operators, $\hat{S}_{m} = \hat{S}_{m,S}^{sp}\otimes\hat{S}_{m,I}^{sp}$.

Lastly, the operator $\hat{U}$ that implements the entire evolution through a DTQW of $M$ round trips can be expressed as $\hat{U} = \hat{U}_M\ldots\hat{U}_1$, with $\hat{U}_m=\hat{S}_m\hat{C}_m$.
Additionally, it is also possible to imprint time-bin-dependent phases on the quantum state before DTQW.
The associated unitary transformation is given by
\begin{align}
    \hat{V}=\sum_{i,j=1}^N e^{i(\varphi_i + \varphi_j)}\ket{ij}\!\bra{ij}.
\end{align}
For illustration, we compute the photon-detection-probability distribution of a two-bin initial state after a simple DTQW that interferes the two populated bins at indices $q$ and $r$ with $q < r$ in analogy to Fig.~\ref{fig:compound_all}.
The initial state after injection into the loops is given by
\begin{align}
    \begin{split}
        \hat{\rho} = &\sum_{\substack{i,j,k,l\\\in\{q,r\}}}^N \rho_{ij}^{kl} \ket{ij}\!\bra{kl}\\
        \xrightarrow[\mathrm{inject}]{\mathrm{loop}} &\sum_{\substack{i,j,k,l\\\in\{q,r\}}}^N \rho_{ij}^{kl} \ket{iSjS}\!\bra{kSlS},  
    \end{split}
\end{align}
with the density-matrix elements $\rho_{ij}^{kl} \coloneqq \bra{ij}\hat{\rho}\ket{kl}$.
To obtain the time-bin probability distribution of the state after the DTQW, e.g., $\bra{rSrS}\hat{U}\hat{\rho}\hat{U}^{\dagger}\ket{rSrS}$, we can compute the backward propagation of the projected state through the quantum walk.
The problem can be simplified by making use of the product form of $\hat{U}$ and computing the single-photon evolution,
\begin{align}
    \begin{split}
        (\hat{U}^{sp})^\dagger\ket{rS} &= (\hat{U}_1^{sp})^\dagger\ldots(\hat{U}_{M-1}^{sp})^\dagger \, \frac {1}{\sqrt{2}} \left(\ket{rS} - i \ket{rL}\right)\\
        &=(\hat{U}_1^{sp})^\dagger \, \frac {1}{\sqrt{2}} \left(\ket{rS} - i \ket{q + 1,L}\right)\\
        &=\frac{1}{\sqrt{2}} \left(\ket{rS} - \ket{qS} \right)
    \end{split}
\end{align}

In this way, the expectation value can be restated in terms of the initial-state coherences.
The probabilities of the four possible detection events then read
\begin{align}
    \begin{split}
        &\bra{rSrS}\hat{U}\hat{\rho}\hat{U}^{\dagger}\ket{rSrS} =\frac14\big[\rho_{qq}^{qq}+\rho_{rr}^{rr}+\rho_{qr}^{qr}+\rho_{rq}^{rq}+\\
        &\qquad(-\rho_{qq}^{qr} - \rho_{qq}^{rq} + \rho_{qq}^{rr} + \rho_{qr}^{rq}
        - \rho_{qr}^{rr} - \rho_{rq}^{rr} + \mathrm{h.c.})\big],
    \end{split}\\
    \begin{split}
        &\bra{r'Lr'L} \hat{U} \hat{\rho}  \hat{U} ^ {\dagger}\ket{r'Lr'L} =\frac14\big[\rho_{qq}^{qq}+\rho_{rr}^{rr}+\rho_{qr}^{qr}+\rho_{rq}^{rq}+\\
        &\qquad(\rho_{qq}^{qr} + \rho_{qq}^{rq} + \rho_{qq}^{rr} + \rho_{qr}^{rq}
        + \rho_{qr}^{rr} + \rho_{rq}^{rr} + \mathrm{h.c.})\big],
    \end{split}\\
    \begin{split}
        &\bra{rSr'L}\hat{U}\hat{\rho}\hat{U}^{\dagger}\ket{rSr'L} =\frac14\big[\rho_{qq}^{qq}+\rho_{rr}^{rr}+\rho_{qr}^{qr}+\rho_{rq}^{rq}+\\
        &\qquad(\rho_{qq}^{qr} - \rho_{qq}^{rq} - \rho_{qq}^{rr} - \rho_{qr}^{rq}
        - \rho_{qr}^{rr} + \rho_{rq}^{rr} + \mathrm{h.c.})\big],
    \end{split}\\
    \begin{split}
        &\bra{r'LrS}\hat{U}\hat{\rho}\hat{U}^{\dagger}\ket{r'LrS} =\frac14\big[\rho_{qq}^{qq}+\rho_{rr}^{rr}+\rho_{qr}^{qr}+\rho_{rq}^{rq}+\\
        &\qquad(-\rho_{qq}^{qr} + \rho_{qq}^{rq} - \rho_{qq}^{rr} - \rho_{qr}^{rq}
        + \rho_{qr}^{rr} - \rho_{rq}^{rr} + \mathrm{h.c.})\big],
    \end{split}
\end{align}
where we have used the shorthand $r' = r + 1$.
It is straightforward to see that the difference of coincidence and non-coincidence probabilities $p_c - p_{nc}$ can be expressed as
\begin{align}
    \begin{split}
        p_c - p_{nc}=& \bra{rSrS}\hat{U}\hat{\rho}\,\hat{U}^{\dagger}\ket{rSrS}\\
        &+ \bra{r'Lr'L}\hat{U}\hat{\rho}\,\hat{U}^{\dagger}\ket{r'Lr'L}\\
        &-\bra{rSr'L}\hat{U}\hat{\rho}\hat{U}^{\dagger}\ket{rSr'L}\\
        &-\bra{r'LrS}\hat{U}\hat{\rho}\hat{U}^{\dagger}\ket{r'LrS}\\
        =&2\operatorname{Re}\left[\bra{qq}\hat{\rho}\ket{rr} + \bra{qr}\hat{\rho}\ket{rq}\right],\label{eq:coincidence}
    \end{split}
\end{align}
which is the result given in Eq.~\eqref{eq:compound}.

\section{Comprehensive DTQW scheme}
\label{app:scheme}

\begin{figure}
    \centering
    \includegraphics[width=\linewidth]{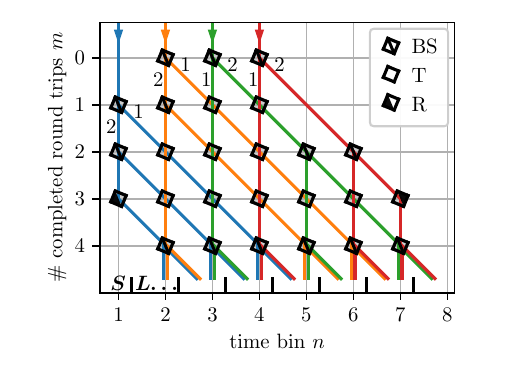}
    \caption{A diagrammatic illustration of the single comprehensive DTQW configuration for $N=4$.
All six two-bin interference patterns are realized for $M=4$.
For the first round trip, the central coupler is configured as a beam splitter with a $1\!:\!2$-splitting ratio, as indicated by the annotations at the two beam-splitter outputs.}
    \label{fig:comprehensive_N4}
\end{figure}

In Sec.~\ref{sec:compound}, we introduce a set of DTQW settings that realize pairwise path interference between all time bins.
It is also possible to generate all pairings of two bins in a single DTQW run.
We have illustrated the corresponding configuration for $N=4$ in Fig.~\ref{fig:comprehensive_N4}.

To achieve symmetric time-bin trajectories for all outcomes in a minimal number of round trips, we employ unbalanced $1\!:\!2$ beam splitters during the first two round trips, corresponding to beam-splitter angles of $\theta=\arccos{\frac{1}{\sqrt{3}}}$ and $\theta=\arccos{\sqrt{\frac{2}{3}}}$, respectively.
This ansatz covers all relevant two-bin interference measurements for the compound scheme,

\begin{align}
    \begin{split}
       \Delta p_c(i,j) = \frac{2}{(N-1)^2}\operatorname{Re}\left[\bra{ii}\hat{\rho}\ket{jj} + \bra{ij}\hat{\rho}\ket{ji}\right],\label{eq:comprehensive}
     \end{split}
 \end{align}
where the prefactor has changed to account for the larger number of outputs.
Moreover, this configuration also allows for access to further quantum coherences.
By measuring the probabilities of detecting the two photons in different DTQW branches $r$ and $q$ that share one common connected time bin of origin $k$, one can obtain
\begin{align}
    \begin{split}
       \Delta p_c(i,j,k) \coloneqq& \bra{rSqS}\hat{U}\hat{\rho}\,\hat{U}^{\dagger}\ket{rSqS}\\
        &+ \bra{r'Lq'L}\hat{U}\hat{\rho}\,\hat{U}^{\dagger}\ket{r'Lq'L}\\
        &-\bra{rSq'L}\hat{U}\hat{\rho}\hat{U}^{\dagger}\ket{rSq'L}\\
        &-\bra{r'LqS}\hat{U}\hat{\rho}\hat{U}^{\dagger}\ket{r'LqS}\\
        =&\frac{2s_k}{(N-1)^2}\operatorname{Re}\left[\bra{kk}\hat{\rho}\ket{ij} + \bra{ik}\hat{\rho}\ket{kj}\right],\label{eq:comprehensive_shared_color}
     \end{split}
\end{align}
with the signal photon detected in bin $r$ and traceable to bins $i$ and $k$ and the idler detected in bin $q$ and traceable to bins $j$ and $k$ and $r'=r+1$ as before.
The factor $s_k \in\{+1,-1\}$ depends on the trajectories of bin $k$ in the two branches.
If the number of loop switches is identical for the short-loop exit port for both branches, e.g., for all branches that share a common blue line in Fig.~\ref{fig:comprehensive_N4}, then $s_k=1$.
If the number is unequal, then $s_k=-1$, as can be seen, for example, for the orange line with $r=2$ and $q=5$.
By combining measurements of $\Delta p_c(i,j,k)$ taken under different initial-state phase profiles, the real and imaginary parts of both coherences can be determined individually, analogous to Eq.~\eqref{eq:real_imag}.
The three additional phase profiles needed are $\varphi_k = \pi/2$, $\varphi_k = \pi/4$, and $\varphi_j=-\varphi_i=\pi/4$.

Finally, the signal and idler photon can be detected in two DTQW branches that share no common time bin of origin but rather link back to bins $i$ and $j$ for one photon and $k$ and $l$ for the other.
Using the corresponding measured event rates, we can express the detection-probability imbalance as
\begin{align}
    \begin{split}
       -\Delta p_c(i,j,k,l) \coloneqq& -\bra{nSmS}\hat{U}\hat{\rho}\hat{U}^{\dagger}\ket{nSmS}\\
        &- \bra{n'Lm'L}\hat{U}\hat{\rho}\hat{U}^{\dagger}\ket{n'Lm'L}\\
        &+\bra{nSm'L}\hat{U}\hat{\rho}\hat{U}^{\dagger}\ket{nSm'L}\\
        &+\bra{n'LmS}\hat{U}\hat{\rho}\hat{U}^{\dagger}\ket{n'LmS}\\
        =&\frac{2}{(N-1)^2}\operatorname{Re}\left[\bra{ik}\hat{\rho}\ket{jl} + \bra{il}\hat{\rho}\ket{jk}\right].\label{eq:comprehensive_different_colors}
     \end{split}
\end{align}
Similarly as before, each individual real and imaginary part of both coherences is accessible by combining a phase-free run with three different initial-state phase imprints given by $\varphi_i = \varphi_k = \pi/2$,  $\varphi_i = \varphi_k = \pi/4$, and $\varphi_i = -\varphi_k = \pi/4$.

This allows us to explicitly determine the real and imaginary parts of all coherences of the form $\bra{ii}\hat{\rho}\ket{jj}$, $\bra{kk}\hat{\rho}\ket{ij}$, $\bra{ik}\hat{\rho}\ket{kj}$, and $\bra{ik}\hat{\rho}\ket{jl}$, where $i,j,k,l$ are all pairwise different and each individual real or imaginary part can be extracted by combining different phase imprints.

Furthermore, by considering the imbalance between the exit ports connected to the short and the long loop, i.e., \hbox{$\sim \bra{nSmS} \hat{U} \hat{\rho} \hat{U}^{\dagger} \ket{nSmS} -\bra{n'Lm'L} \hat{U} \hat{\rho} \hat{U}^{\dagger} \ket{n'Lm'L}$} and \hbox{$\sim \!\bra{nSm'L} \hat{U} \hat{\rho} \hat{U}^{\dagger} \ket{nSm'L} \! - \! \bra{n'LmS} \hat{U} \hat{\rho} \hat{U}^{\dagger} \ket{n'LmS}$}\, one can access most of the remaining coherences as well.
However, due to symmetry-induced phase cancellation, only the real part of the coherences of the form $\bra{ij}\hat{\rho}\ket{ji}$ is accessible due to the fact that the applied DTQW acts identically on the signal and idler photons.
This means that even though any quantum state can be classified to a large degree, full quantum state tomography will remain out of reach for the given experimental capabilities.

\section{Compound DTQW configuration for \textit{N}=8}
\label{app:compound}

\begin{figure*}[ht]
    \centering
    \includegraphics[width=0.7\linewidth]{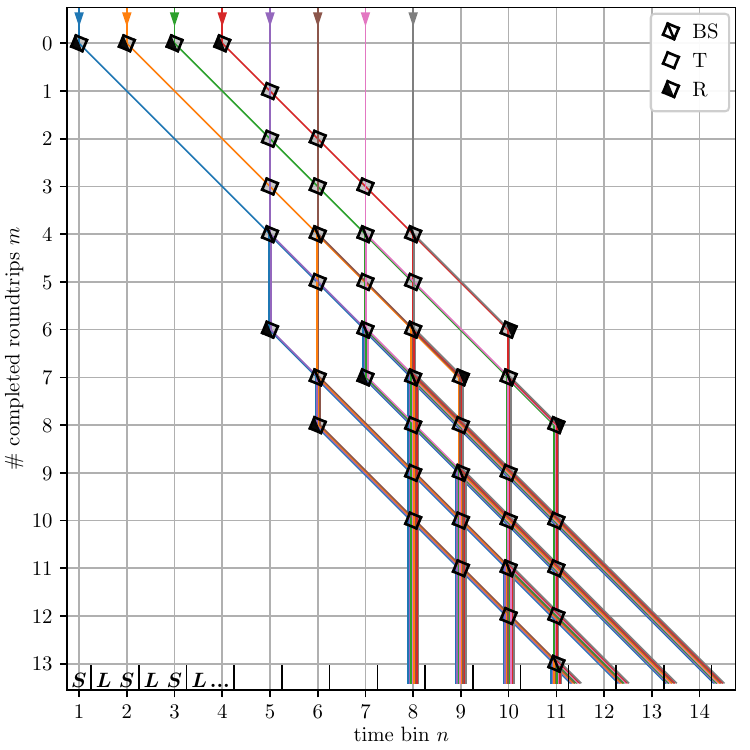}
    \caption{An illustration of the single-setup DTQW configuration for $N=8$.
The scheme interferes all initial-state bins in each of the $N$ output bins.}
    \label{fig:single_setup_N8}
\end{figure*}

In Fig.~\ref{fig:single_setup_N8} we give the illustration of the DTQW configuration described in Sec.~\ref{sec:single-setting} for $N=8$.
The configuration for $N=4$ appears again between $m=4$ and $m=9$ in the central branch of the DTQW.
The earlier and later bins with partial information are interfered within their branches in $m=7$ in time bins $n=6$ and $n=10$.
Finally, these bins interfere in $m=11$ and $m=13$, so that thereafter all bins carry full interference information.

\section{Multi-SPDC}
\label{app:multi-spdc}

Here we give more details about the treatment of states with two SPDC pairs, i.e., four-photon states with two indistinguishable signal photons and two indistinguishable idler photons.
First, the coin operator has to be adapted to correctly act on two incident photons of the same mode.
Here, one has to differentiate between the three different incoming states $\ket{SS}=(\hat{a}_S^\dagger)^2\ket{0}/\sqrt{2}$, $\ket{SL}=\hat{a}_S^\dagger\hat{a}_L^\dagger\ket{0}$, and $\ket{LL}=(\hat{a}_L^\dagger)^2\ket{0}/\sqrt{2}$.
The central coupler acts on the creation operators as 
\begin{align}
   \begin{pmatrix}
        \hat{a}_S^\dagger\\
        \hat{a}_L^\dagger
    \end{pmatrix}
    \rightarrow
    \begin{pmatrix}
        \cos\theta & i\sin \theta \\
        i\sin \theta & \cos\theta
    \end{pmatrix}
    \begin{pmatrix}
        \hat{a}_S^\dagger\\
        \hat{a}_L^\dagger
    \end{pmatrix}
\end{align}
or, equivalently, on the incoming states as
\begin{align}
    \begin{split}
        \hat{C} = &\cos^2\theta\left(\ket{SS}\bra{SS} + \ket{LL}\bra{LL} \right)\\
        - &\sin^2\theta\left(\ket{SS}\bra{LL} + \ket{LL}\bra{SS}\right)\\
        + &\left(\cos^2\theta - \sin^2\theta\right)\ket{SL}\bra{SL}\\
        +&i\sqrt2\sin\theta\cos\theta\left(\ket{SL}\bra{SS} + \ket{SL}\bra{LL} + \mathrm{h.c.}\right),    
    \end{split}
\end{align}
having omitted the time-bin index $n$ and the round-trip index $m$.
In particular, for a balanced beam splitter with $\theta=\pi/4$, the weight of the $\ket{SL}\bra{SL}$ term vanishes, which is known as the Hong-Ou-Mandel effect \cite{hong_measurement_1987}.
As we show below, we utilize this effect in the formulation of our fidelity bound.
First, we construct the part of the bound based on the DTQW scheme introduced in Sec.~\ref{sec:multi-spdc}.
All relevant measurements are taken at the central four-photon interference.

By combining the measured probabilities to detect all four photons in the same branch ($p_{FC}^{\alpha}$), both idler photons in the same branch and both signal photons in the other ($p_{SC}^{\alpha}$), and finally one signal-idler pair in one branch and one pair in the other ($p_{PC}^{\alpha}$), we can obtain the following expression,
\begin{widetext}
\begin{align}
        \begin{split}
            8\sum_\alpha p_{FC}^{\alpha} - p_{SC}^{\alpha} + 2p_{PC}^{\alpha}=
            \frac 34&\sum_{i, j}^N \bra{iiii}\hat{\rho}\ket{jjjj}
            +\sum_{\substack{i\leq j\\ k\leq l}}^N \bra{ijij} \hat{\rho} \ket{klkl} \mathbf{1}_{l+j-k-i>0}\\
            +&\sum_{\substack{m\leq n\\o\leq p}}^N\sum_{\substack{i\leq j\\ k\leq l}}^N  g_{ijmn}^{klop} \bra{ijmn} \hat{\rho} \ket{klop}\mathbf{1}_{|m-i| + |n-j|+|o-k| + |p-l|>0},
        \end{split}\label{eq:multi-spdc-4bin}
    \end{align}
    where the index $\alpha\in\{a, b, c\}$ relates to the corresponding panel in Fig.~\ref{fig:compound-multi-spdc}.
In the above equation, we have introduced indicator functions $\mathbf{1}_{x>0} = 1$ for $x>0$ and $\mathbf{1}_{x>0} = 0$ otherwise.
The first indicator function excludes all terms of the form $\bra{iiii}\hat{\rho}\ket{jjjj}$ from the second sum on the RHS of the equation.
Similarly, the second indicator function removes all terms of the form $\bra{ijij}\hat{\rho}\ket{klkl}$ from the last sum.
    
All coherences that contribute to the fidelity to the maximally entangled reference state $\Phi_{\mathrm{MES}}$ given in Eq.\eqref{eq:fidelity-multi-spdc} appear on the RHS of Eq.~\eqref{eq:multi-spdc-4bin}, but not with the correct relative amplitudes.
Specifically, coherences of the form $\bra{iiii}\hat{\rho}\ket{jjjj}$ have a deviating weight of $3/4$.
As stated above, one can utilize the Hong-Ou-Mandel effect in two-photon interference to specifically target coherences with fully coincident signal and idler photons.
Hence, we propose to measure the probability to detect a signal-idler pair in each output of the final beam splitter connecting bins $i$ and $j$ ($p_{PC}(i,j)$) in the compound two-bin interference scheme introduced in Sec.~\ref{sec:compound} (see Fig.~\ref{fig:compound_all}).
The sum over all combinations of bins can be expressed as
    \begin{align}
       \begin{split}
            \sum_{i<j}^N p_{PC}(i,j) = \frac 14&\sum_{i\neq j}^N \bra{iiii}\hat{\rho}\ket{jjjj} + \frac 34\sum_{i}^N \bra{iiii}\hat{\rho}\ket{iiii}
            +\sum_{\substack{i, j\\ k, l}}^N d_{iijj}^{kkll} \bra{iijj} \hat{\rho} \ket{kkll}\mathbf{1}_{|j-i| + |l-k| >0},\label{eq:multi-spdc-HOM}
        \end{split}
    \end{align}
     with the indicator function removing all terms of the kind $\bra{iiii}\hat{\rho}\ket{jjjj}$, as before.
By combining Eqs.~\eqref{eq:multi-spdc-4bin} and~\eqref{eq:multi-spdc-HOM}, one can equalize the weights of all coherences, but overcompensates for fully correlated time-bin populations of the form $\bra{iiii}\hat{\rho}\ket{iiii}$.
However, as these are known exactly through time-of-arrival measurements directly after the SPDC, this can be remedied by subtracting the corresponding terms.
To conclude, adding all these results yields 
    
    \begin{align}
        \begin{split}
            \mathcal{C}=8\sum_\alpha p_{FC}^{\alpha} - p_{SC}^{\alpha} + 2p_{PC}^{\alpha} -\frac 12&\sum_{i}^N \bra{iiii}\hat{\rho}\ket{iiii} + \sum_{i<j}^N  p_{PC}(i,j)  =\\
            &\sum_{\substack{i\leq j\\ k\leq l}}^N \bra{ijij} \hat{\rho} \ket{klkl}
            +\sum_{\substack{m\leq n\\o\leq p}}^N\sum_{\substack{i\leq j\\ k\leq l}}^N c_{ijmn}^{klop} \bra{ijmn} \hat{\rho} \ket{klop}\mathbf{1}_{|m-i| + |n-j|+|o-k| + |p-l|>0},
        \end{split}
    \end{align}
with the first sum on the RHS proportional to the fidelity to $\Phi_{\mathrm{MES}}$ and coefficients $c_{ijmn}^{klop} \coloneqq g_{ijmn}^{klop} + d_{ijmn}^{klop}$.
This is the result shown in Eq.~\eqref{eq:multi-spdc-C}.

A further improvement can be achieved by running the four-bin interference in four different phase profiles.
In each profile, a phase of $\varphi_n=\pi$ is imprinted on one of the four time bins, resulting in
    \begin{align}
        \begin{split}
            2\sum_{n=1}^N\sum_\alpha \big(p_{FC}^{\alpha} - p_{SC}^{\alpha} + 2p_{PC}^{\alpha}\big)_{\big|\varphi_n=\pi}=
            \frac 34&\sum_{i, j}^N \bra{iiii}\hat{\rho}\ket{jjjj}
            +\sum_{\substack{i\leq j\\ k\leq l}}^N \bra{ijij} \hat{\rho} \ket{klkl} \mathbf{1}_{l+j-k-i>0}\\
            +&\sum_{\substack{m\leq n\\o\leq p}}^N\sum_{\substack{i\leq j\\ k\leq l}}^N  \tilde{c}_{ijmn}^{klop} \bra{ijmn} \hat{\rho} \ket{klop}\mathbf{1}_{|m-i| + |n-j|+|o-k| + |p-l|>0},
        \end{split}\label{eq:multi-spdc-4bin-phase}
    \end{align}
with modified weights $\tilde{c}_{ijmn}^{klop}$.
Importantly, $\tilde{c}_{ijij}^{klkm}=\tilde{c}_{ijij}^{lkmk} = 0$, for $l\neq m$, i.e., the weights of all coherences between macroscopically populated states and those including two uncorrelated photons vanish.
This corresponds to the simple dephasing model $\hat{\rho} = (1-\alpha)\ket{\Psi}\!\bra{\Psi} + \alpha\hat{\rho}_{\mathrm{deph.}}$ and $\alpha\ll1$, where these terms are the leading error terms which cannot be removed in postselection.
This can be seen from the bound in Eq.~\eqref{eq:Zij}, as $|\bra{ijij}\hat{\rho}\ket{lkmk}|\leq \sqrt{\bra{ijij}\hat{\rho}\ket{ijij}\bra{lkmk}\hat{\rho}\ket{lkmk}}\propto\sqrt{\alpha}$, whereas all other coherences that do not contribute to the fidelity have $|\bra{ijkl}\hat{\rho}\ket{mnop}|\leq \sqrt{\bra{ijkl}\hat{\rho}\ket{ijkl}\bra{mnop}\hat{\rho}\ket{mnop}}\propto\alpha$.
Thus, removing these coherences from the last sum in Eq.~\eqref{eq:multi-spdc-4bin-phase} significantly improves the bound tightness, as described in Sec.~\ref{sec:multi-spdc}.

\end{widetext}
\bibliography{mybib}

\end{document}